\documentclass{PoS}

\usepackage{epsfig}
\usepackage{amsfonts}
\usepackage{amssymb} 			
\usepackage{amsmath} 			
\usepackage{verbatim}		 	
\usepackage{psfrag} 
\usepackage{subfigure} 

\newcommand{\interi} {\( \angle \kern -.5em\mathrm{Z} \)} 		
\newcommand{\reali} {\( \textsf {I}\kern -.15em\mathrm{R} \)} 	

\newcommand{\be}{\begin{equation}}
\newcommand{\ee}{\end{equation}}

\newcommand{\s}{\textsl}

\newcommand{\Turns}[0]{\resizebox{1.7\width}{1.3\height}{$\;\leadsto\;$}}

\newcommand{\hgamma}[0]{{\resizebox{!}{5.36pt}{$\gamma$}}}

\newcommand{\hpi}[0]{{\resizebox{!}{5.36pt}{$\pi$}}}

\newcommand{\lmu}[0]{{\resizebox{!}{3.91pt}{$\mu$}}}

\newcommand{\bbar}[1]{\,\overline{\!#1}}

\newcommand{\gGamma}{\Gamma_{\!2\phantom{\hat{i}}}\!\!}

\newcommand{\pslash}{\diagup\!\!\!\!\!\!\!p}

\newcommand{\Gslash}{\raisebox{0.080em}{$\diagup$}\hspace{-.76em}\Gamma}

\newcommand{\Inv}[2]{\mathbb{S}^{\!(#1)}_{\phantom{1\!\!\!}#2}\!}
\newcommand{\inv}{\mathbb{S}}
\newcommand{\InV}[2]{\mathbb{V}^{(#1)}_{\phantom{1\!\!\!}\!\!#2}\!}

\newcommand{\1}{{\sf 1\hspace*{-0.3ex}%
    \rule{0.15ex}{1.5ex}
    \rule{.25ex}{.1122ex}
    \hspace*{0.3ex}%
    }}

\newcommand{\HAT}[1]{{\hat{#1\,}\!}}

\newcommand{\n}[1]{\ensuremath{\mathrm{#1}}}

\PoS{PoS(LAT2005)108}

\title{Fermionic observables in Numerical Stochastic Perturbation Theory }

\ShortTitle{Fermionic observables in Numerical Stochastic Perturbation Theory }

\author{\speaker{Vincenzo Miccio}\thanks{Current address: INFN Milano}\\
        University of Parma and INFN\\
        E-mail: \email{vincenzo.miccio@mib.infn.it}}

\author{Francesco Di Renzo\\
        University of Parma and INFN\\
        E-mail: \email{direnzo@parma.infn.it}}

\author{Andrea Mantovi\\
        University of Parma and INFN\\
        E-mail: \email{mantovi@parma.infn.it}}

\author{Christian Torrero\\
        University of Parma and INFN\\
        E-mail: \email{torrero@parma.infn.it}}

\author{Luigi Scorzato\\
				Humboldt Universit\"at\\
        E-mail: \email{luigi.scorzato@physik.hu-berlin.de}}

\abstract{We present technical details of fermionic observables computations in NSPT. In particular we discuss the construction of composite operators starting from the inverse Dirac operator building block, the subtraction of UV divergences and the treatment of irrelevant contributions in extracting the continuum limit.}

\FullConference{XXIIIrd International Symposium on Lattice Field Theory\\
25-30 July 2005\\
Trinity College, Dublin, Ireland}

\begin{document}

\section*{Introduction}
Motivations and a full account of Numerical Stochastic Perturbation Theory (NSPT) method for Lattice Gauge Theories, with special regard to dynamical fermions, can be found in \cite{FranzGigi2004} and references therein. In another contribution in this conference \cite{FranzLattice2005} we present 
our recent results. Here we want to stress some technical aspects, dealing in particular with fermionic observables computations.

\section{Langevin evolution}
According to the {\it Stochastic Quantization} approach
, we randomly sample the phase-space of the generic field theory with action $S[\phi]$, according to the Langevin equation:
\begin{equation}
\frac{\partial\phi(x,t)}{\partial t}=-\frac{S[\phi]}{\partial\phi(x,t)}+\eta(x,t) \;;
\end{equation}
$t$ can be regarded as a new, non-physical, stochastic time. This sampling is such that the average over the gaussian noise $\eta$ leads, for infinite time evolution, to 
the same expectation values of Feynman's path-integrals:
\begin{equation}
\left\langle \mathcal{O}[\phi_\eta(t)]\right\rangle_\eta 
\mathop{-\hspace{-.2cm}-\hspace{-.2cm}-\hspace{-.2cm}-\hspace{-.2cm}\longrightarrow}_{t\rightarrow\infty}
\frac{1}{Z}\hspace{-.1cm}\int\hspace{-.1cm}\mathcal{D}[\phi]\mathcal{O}[\phi]\exp\{-S[\phi]\} \;.
\end{equation}
Perturbation theory is performed by thinking of the field as a formal power series in the coupling constant $\lambda$:
\begin{equation}
\phi_\eta(x;t) \Turns\!\!\sum_n\lambda^n\phi_\eta^{(n)}(x;t)\;.
\end{equation}
Plugging it in the Langevin equation results in a \s{hierarchical} system of differential equations, one for each perturbative term $\phi_\eta^{(i)}$ of the series. 
We solve it numerically {\it via} discretization of the stochastic time $t=n\tau$.
Every dynamical variable has to be replaced by its perturbative (truncated) expansion 
\begin{equation}{\textstyle
\mathsf{A}\Turns\!\!\{\mathsf{A}^{(n)}\}_{n\,\leq\,\bbar{n}}\;,
}\end{equation} 
that is by a \s{collection} of dynamical variables. Consistently, every algebraic operation has to be performed `order by order': for additions and scalar-multiplications this simply means to `vectorize' them,
\begin{eqnarray}\label{eq:SumOrdByOrd}
\left(a\;\mathsf{A}\right)^{(n)}&=&a\;\mathsf{A}^{(n)}\\
\left(\mathsf{A}+\mathsf{B}\right)^{(n)}&=&\;\mathsf{A}^{(n)}+\mathsf{B}^{(n)}\hspace{5em} n=0,1,\ldots,\bbar{n}\;,
\end{eqnarray}
while ordinary multiplications have to be replaced by the (truncated) Cauchy product:
\begin{equation}\label{eq:OrdByOrd}
\left(\mathsf{A}\;\mathsf{B}\right)^{(n)}=\sum_{j=0}^n\mathsf{A}^{(j)}\mathsf{B}^{(n-j)}\hspace{6.5em} n=0,1,\ldots,\bbar{n}\;.
\end{equation}
The need for a collection of variables in the place of a single one makes simulations very \s{memory demanding}. Moreover, since the computational effort of each basic operation is increased by the order-by-order mechanism, the higher loop one wants to compute, the more expensive simulations are. 
In spite of this, two remarks are in order. First, there are cases where one can take advantage of the perturbative approach (see Sec.\ref{sec:invM})
. The second remark is related to parallel-computing, where inter-nodes communications represent the bottleneck for a true scalability of performances (and hence of the overall computational power). The communications cost is proportional to the amount of data to be transferred, i.e. perturbative simulations communication time increases linearly with the maximum perturbative order $\bbar{n}$. On the contrary, floating point operations increase quadratically with $\bbar{n}$, namely as $\bbar{n}(\bbar{n}-1)/2$. So, as $\bbar{n}$ increases, one improves the ratio between the computation and the communication times.

\section{Inverting the Dirac operator: unquenched dynamics and fermionic observables}
\label{sec:invM}
One of the main advantage of NSPT is that it is surprisingly cheap to invert the Dirac operator $M$. This is needed both for 
fermionic observables computations and \textit{unquenched} dynamics. 

During dynamics, one has to compute the drift term of the Langevin equation
\be\label{eq:fermionicdrift}
\nabla_{\!\!a}S_F^{{}^\n{eff}} = -n_f\,\n{Tr}\bigl[\nabla_{\!\!a}M\;\;M^{-1}\bigr]
\ee
for the fermionic action $S_F^{{}^\n{eff}}=-n_f\,\n{Tr}\;\n{log}M$. The trick is to recover \eqref{eq:fermionicdrift} as an average over a gaussian noise $\xi$
\be
\nabla_{\!\!a}S_F^{{}^\n{eff}} =  -
		n_f\,\bigl[\;\:\xi_k^\dagger\,
			(\nabla_{\!\!a}M)_{kl}
			\,\bigl(M^{-1}\bigr)_{lj}\;\xi_j
		\:\;\bigr]\;,
\ee
so that one is left with the problem of inverting $M$ over a source: $M_{\!_{\scriptstyle kj}}\Psi_j = \xi_k$. In a perturbative context, higher orders $\Psi$ can be recursively computed starting from the inversion of the $0$-order only. The key point is that such a $0$-order inversion is the tree-level (lattice) Feynmann propagator, which does not depend on the field configuration and is diagonal in momentum space. Hence, all the non-locality character of the problem is reduced to an iterative application (back and forth) of a (fast) Fourier transform (for details, see \cite{FranzGigi2004}).

The same procedure is needed for the measurement of the quark propagator. Now we do not have to plug a noise $\xi$ but a $\delta$-source for each element $i$ of the propagator\footnote{It is worthwhile to note that we measure propagator in momentum space, as opposed to the non-perturbative practice. So in the $\delta$-source, $i$ is actually a momentum space index.} we are interested in: $M_{\!_{\scriptstyle kj}}\Psi_j = \delta_k^i$. A slight generalization of the same procedure can be used to compute fermion currents. The generic bilinear operator reads:
\be
\delta\:(M^{-1}\,G\;M^{-1})\:\delta\;,
\ee
where $G$ is a generic Dirac matrix: $G=\1,\,\,\gamma_5,\,\,\gamma_\mu,\,\,\gamma_\mu\gamma_5$ and $\sigma_{\mu\nu}\propto[\gamma_\mu,\gamma_\nu]$. It can be obtained by means of suitable scalar products involving again the inverse of Dirac operator:
\be
(\delta\:M^{-1})\;G\;(M^{-1}\delta)\;.
\ee
This strategy aims at saving CPU-time, but, as a drawback, it requires a lot of memory. Even if you are interested in (a small amount of) diagonal entry (in momentum space) of the current operators, you have to sum over the inner dummy indices, which, as said, extend all over the lattice volume.
\footnote{As a matter of fact, our APEmille machine does not have enough memory to perform such a computation on-the-fly, and so it has to break up the mechanism (carrying out some computations twice). On the other hand, instead, the memory equipment of a PC-cluster can do the job.}

\section{Exploiting the hypercubic symmetry to guide the continuum limit}
In order to fix ideas, let us focus on the two points vertex function $\gGamma$.
From a general point of view, with respect to Dirac-spin space, it can be decomposed into its component along the identity and the gamma matrices:
\be\label{gamma2}
\gGamma 	\;\;=\;\; \Gamma_{\!\n{id}} + i\Gslash 
			\;\;=\;\; \Gamma_{\!\n{id}}\,\1 +
						i\,{\textstyle\sum\limits_{\,\mu\;}}\,\hgamma_\mu\Gamma_{\!\mu}\;.
\ee
With respect to space-time, each component $\Gamma_{\!\mu}$ and $\Gamma_{\!\n{id}}$ can be (function of) Lorentz-Euclidean invariant quantities only.
In the continuum $\n{O}(4)$-symmetry case $p^2$ is the only scalar quantity at one's disposal 
and $p_\mu$ the only vector quantity. 
So \eqref{gamma2} would translate in%
\begin{subequations}\label{eq:continuumGammaDependence}\begin{eqnarray}
&&\hspace{-1.8em}\Gamma_{\!\n{id}}\sim\Gamma_{\!\n{id}}(p^2,m) \\
&&\hspace{-1.8em}\Gamma_{\!\mu}\sim p_\mu \:\overline{\Gamma}(p^2,m)
	\hspace{3em}\n{i.e.}
	\hspace{3em}\Gslash\sim\pslash\;\overline{\Gamma}(p^2,m)\;.
\end{eqnarray}\end{subequations}
On the lattice one has to take into account the smaller hypercubic symmetry $W_4$, the crystallographic group of the discrete $\hpi/2$ rotations on the six lattice planes of a 4-d
hypercubic lattice onto itself, with the addition of the reflections. Less symmetry means less constraints, and so more assortment. Within the scalar sector, one has to face combinations of even powers of the momentum components:%
\begin{subequations}
\be
\Inv{2}{}(\HAT{p})=\sum_{\!\mu}\HAT{p}_\mu^2
\ee
\be\label{eq:o4invariants}
\Inv{4}{1}(\HAT{p})=\sum\limits_{\mu\,} \HAT{p}_\mu^4\;\;\;\;,\;\;\;\;\;\;
\Inv{4}{2}(\HAT{p})=\sum\limits_{\mu\,}
	\sum\limits_{\nu\ne\mu}\HAT{p}_\mu^2\HAT{p}_\nu^2\;\;\;\;,
\ee
\be
\Inv{6}{1}(\HAT{p})=
	\sum\limits_{\mu\,}
		\HAT{p}_\mu^6\;\;\;\;,\;\;\;\;\;\;
\Inv{6}{2}(\HAT{p})=
	\sum\limits_{\mu\,}
		\sum\limits_{\nu\ne\mu}
			\HAT{p}_\mu^4\HAT{p}_\nu^2\;\;\;\;,\;\;\;\;\;\;
\Inv{6}{3}(\HAT{p})=
	\sum\limits_{\mu\,}
		\sum\limits_{\nu\ne\mu}
			\sum\limits_{\substack{\rho\ne\mu\\\rho\ne\nu}}
				\HAT{p}_\mu^2\HAT{p}_\nu^2\HAT{p}_\rho^2	
\ee
\end{subequations}
and so on.\footnote{~%
Strictly speaking, the ones listed above represent only a particular \s{base} for the generic invariants of a given order. The equally invariant object $\bigl(\sum_\mu\!p_\mu^2\bigr)^2$, for example, can be obtained by means of a suitable combination of the \eqref{eq:o4invariants}'s.} 
Scalar invariants would suffice if we are interested in $\Gamma_{\!\n{id}}$ only, since, as said, itself is a scalar. 
$\Gamma_{\!\mu}$, in turns, has to transform as a vector under the $W_4$ symmetry, so that
one has to take into account any 
odd\footnote{~Even powers are forbidden here --- just likewise we considered, before, no odd powers for the $\inv$ invariants --- because the hypercubic group symmetry does include axes reflections.} powers of 
$\HAT{p}_\mu$: 
\be
\InV{1}{\mu}(\HAT{p})=\HAT{p}_\mu\;\;\;\;,\;\;\;\;\;\;\InV{3}{\mu}(p)=\HAT{p}_\mu^3\;\;\;\;,\;\;\;\;\;\;\InV{5}{\mu}(\HAT{p})=\HAT{p}_\mu^5\;\;\;\;,\;\;\;\;\;\;\ldots
\ee
In the end the expressions \eqref{eq:continuumGammaDependence}'s for the $\gGamma$ translate on the lattice into:
\footnote{From now on we will consider the vanishing bare mass case, $m=0$, 
so that all the mass-dependences can be neglected.}%
\begin{subequations}\label{eq:LatticeGammaDependence}\begin{eqnarray}
&&\hspace{-1.8em}\Gamma_{\!\n{id}}\sim\Gamma_{\!\n{id}}(\,\Inv{2n}{i}\,) \\
&&\hspace{-1.8em}\Gamma_{\!\mu}\sim \Gamma_{\!\mu}(\,\Inv{2n}{i},\InV{2n+1}{\mu}\,)\;.
\end{eqnarray}\end{subequations}

By restoring physical dimensions, each lattice momentum gets its lattice-spacing factor $a$ in front: $\HAT{p}=ap$. So one can recover the desired continuum limit $a\rightarrow0$ just in the limit of vanishing (lattice) momentum $\HAT{p}$. Therefore, a clear procedure for disentangling each quantity from finite lattice-size artifacts relies on a small-momenta expansion, so that only a few of them --- the lower power ones --- have to be really taken into account. As a result, the expansion for the \eqref{eq:LatticeGammaDependence}'s reads:%
\begin{subequations}
\begin{eqnarray}\label{eq:BothInvariant}
\Gamma_{\!\n{id}} &\;=\;& \sigma^{(0)} \;+\;
	\sigma^{(2)}\HAT{p}^2 \;+\; 
	\sum_{i=1}^{2}\sigma^{(4)}_i\Inv{4}{i}\;+\; 
	\sum_{i=1}^{3}\sigma^{(6)}_i\Inv{6}{i}\;+ \ldots\;,\\
\Gamma_{\!\mu} &=& \label{eq:ScalarInvariantsExpansion}
	\;\HAT{p}_\mu\,\;\Bigl[\;
		\vartheta^{(1,0)}\;+\;
		\vartheta^{(1,2)}\HAT{p}^2 \;+\; 
		\sum\nolimits_i
		\vartheta^{(1,4)}_i\Inv{4}{i}\;+\;\ldots\;\Bigr]
		+\nonumber\\
	&+&\;\HAT{p}^3_\mu\;\,\Bigl[\;
		\vartheta^{(3,0)}\;+\;
		\vartheta^{(3,2)}\HAT{p}^2 \;+\; 
		\sum\nolimits_i
		\vartheta^{(3,4)}_i\Inv{4}{i}\;+\;\ldots\;\Bigr]
		+\\
	&+&\;\HAT{p}^5_\lmu\,\;\Bigl[\;
		\vartheta^{(5,0)}\;+\;
		\vartheta^{(5,2)}\HAT{p}^2 \;+\; 
		\sum\nolimits_i
		\vartheta^{(5,4)}_i\Inv{4}{i}\;+\;\ldots\;\Bigr]
		+\;\dots\;.\;\nonumber
\end{eqnarray}
\end{subequations}
From dimensional analysis, the first coefficient of the series represents just the continuum limit value. 
\begin{figure}[!bt]
	\begin{center}

	\psfrag{6.5}[r][r][.58]{6.5}
	\psfrag{7.0}[r][r][.58]{7.0}
	\psfrag{7.5}[r][r][.58]{7.5}
	\psfrag{8.0}[r][r][.58]{8.0}
	\psfrag{8.5}[r][r][.58]{8.5}
	\psfrag{9.0}[r][r][.58]{9.0}
	\psfrag{9.5}[r][r][.58]{9.5}
	\psfrag{10.0}[r][r][.58]{10.0}
	\psfrag{10.5}[r][r][.58]{10.5}
	\psfrag{11.0}[r][r][.58]{11.0}
	\psfrag{11.5}[r][r][.58]{11.5}
	\psfrag{12.0}[r][r][.58]{12.0}

	\psfrag{ms3}[tl][Bl][.9]{ }
	\psfrag{mc3}[cc][cc][.9]{ }
	\psfrag{a2p2}[Bl][Bl][.91]{$a^2\!p^2$}

	\psfrag{0.0}[tc][tc][.58]{0.0}
	\psfrag{0.2}[tc][tc][.58]{0.2}
	\psfrag{0.4}[tc][tc][.58]{0.4}
	\psfrag{0.6}[tc][tc][.58]{0.6}
	\psfrag{0.8}[tc][tc][.58]{0.8}
	\psfrag{1.0}[tc][tc][.58]{1.0}
	\psfrag{1.2}[tc][tc][.58]{1.2}
	\psfrag{1.4}[tc][tc][.58]{1.4}
	\psfrag{1.6}[tc][tc][.58]{1.6}
	\psfrag{1.8}[tc][tc][.58]{}

	\psfrag{0}[tc][tc][.58]{0.0}
	\psfrag{1}[tc][tc][.58]{1.0}

	\psfrag{-0.85}[r][r][.58]{-0.85}
	\psfrag{-0.8}[r][r][.58]{-0.80}
	\psfrag{-0.75}[r][r][.58]{-0.75}
	\psfrag{-0.7}[r][r][.58]{-0.70}
	\psfrag{-0.65}[r][r][.58]{-0.65}
	\psfrag{-0.6}[r][r][.58]{-0.60}

		\newlength{\altezza}
		\setlength{\altezza}{4.6cm}
		\includegraphics[width=1.61803398874989\altezza,height=\altezza]{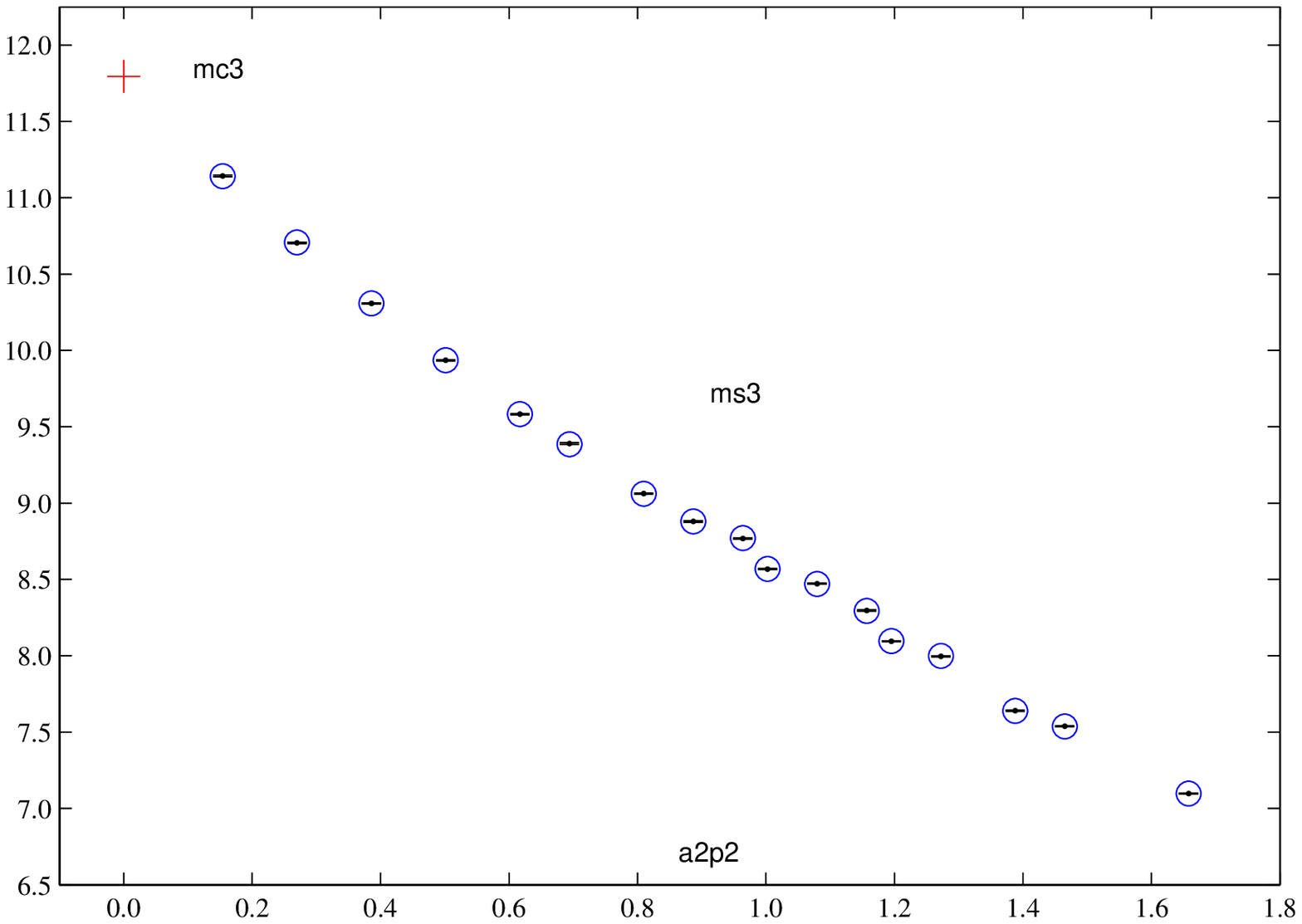}
		\includegraphics[viewport=.8cm .5cm 14cm 11cm,clip,width=1.61803398874989\altezza,height=\altezza]{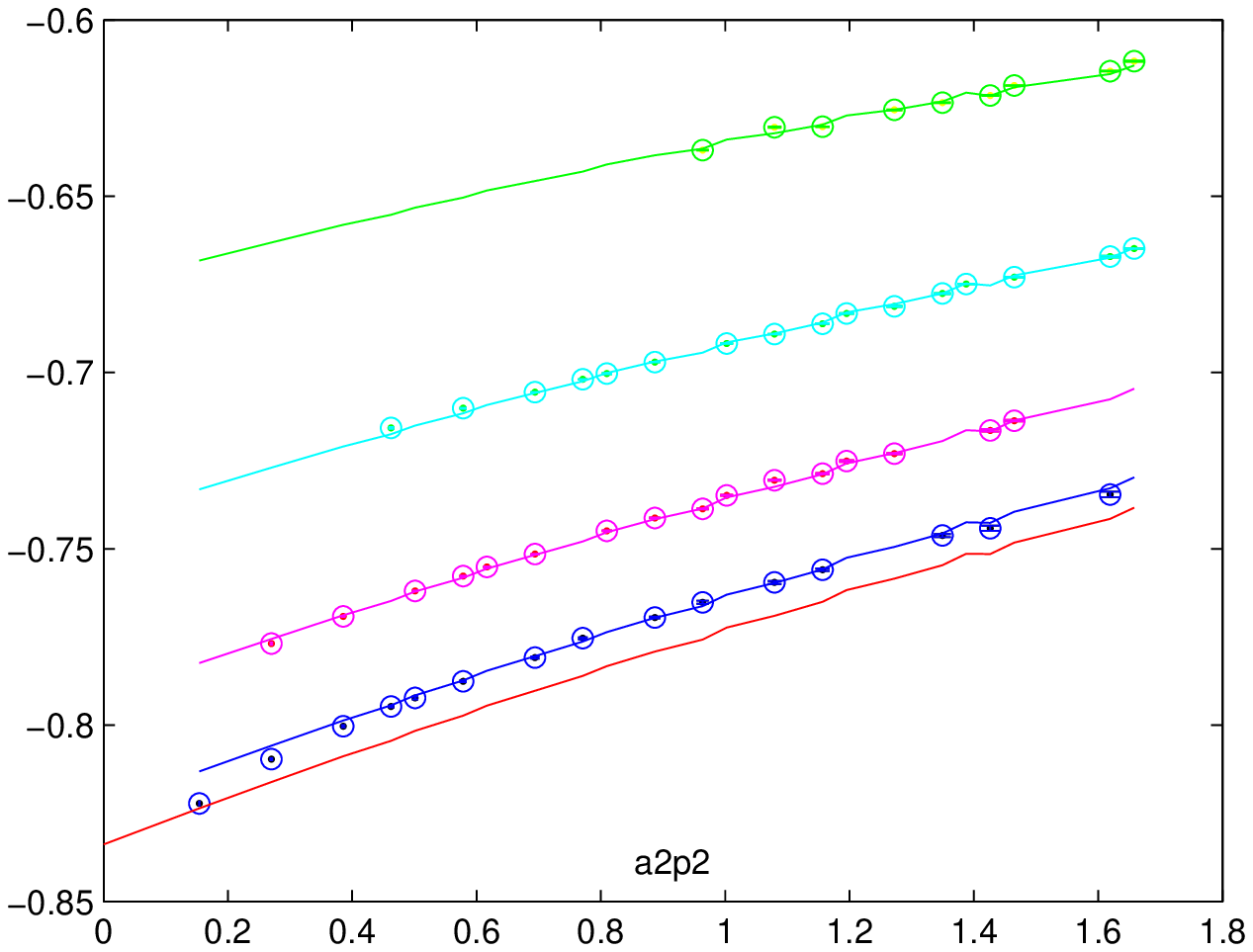}
	\caption{Lattice corrections and continuum extrapolation: example of scalar quantity (on the left) and `vector-like' quantity (on the right). On the left, interpolation fit is represented with circle around numerical data (with errorbars, almost invisible in the picture); on the right, the (more involved) interpolation fit results in the solid lines. See text for details.}
	\label{fig:LatticeCorrection}
\end{center}
\end{figure}
Figure \ref{fig:LatticeCorrection} shows two typical examples. The plot on the left represents measurements of (the third perturbative loop of) the critical mass for Wilson fermions. Since it is a scalar quantity, its behaviour is affected only by scalar lattice corrections, and so the picture is quite simpler with respect to the more general case (see next picture). For practical reasons data are shown versus the leading correction only, the momentum square, but the presence of the other invariants shows up in the jaggedness of the curve (which indeed does not represent statistical fluctuations of the signal). The plot on the right represents measurements of (the first perturbative loop of) the field renormalization constant, which results from the component along $\HAT{p}_\mu$ of $\gGamma$. The situation is quite more involved with respect to the case of a scalar quantity, since here `vector-like' lattice corrections mix up with the scalar ones. So we have not only the jaggedness of the scalar corrections, but points arrange into different levels according to the value of the component of the momentum in the direction along which the gamma-projection was performed. In simpler words: if the considered component of the momentum is equal to just one lattice-momentum-unit, then the associated point will fall on the lower level; if the component is twice, then the associated point will lie on the level above; if the component is three times the lattice-momentum-unit, then the point will lie on the next level; and so on.

\section{Renormalization factors and anomalous dimensions}
In order to obtain the critical mass and the quark wave-function renormalization constant, one has only to get the two points vertex function inverting the fermion propagator. For bilinears, then, one uses the quark propagator (just measured for the same momenta) to amputate the external legs of the current operators. For UV-diverging quantities, care has to be taken while handling the logarithm terms, relying on the (already known) anomalous dimensions.

As an example, let us refer to the scalar current in Landau gauge. The RI' renormalization condition reads:
\be
Z^{-1}_q \,Z_s\; \langle O_s\rangle\rule[-13pt]{.38pt}{18.7pt}_{\:p^2=\lmu^2} = 1
\ee
where $Z_q$ and $Z_s$ are the renormalization factors for the quark wave-function and the scalar current respectively, while $O_s$ is a convenient projection of the scalar current operator as it is measured on the lattice. Expanding everything in the lattice coupling $\beta^{-1}$ up to, for instance, the first order, one gets:
\be
\Bigl(1 - \frac{z_q^{(1)}}{\beta\;} + \ldots\Bigr) 
\Bigl(1+\frac{z_s^{(1)}-\gamma_s^{(1)}\n{log}(p^2)}{\beta\;}  + \ldots\Bigr)
\Bigl(1 - \frac{o_s^{(1)}}{\beta\;} + \ldots\Bigr)\rule[-13pt]{.38pt}{23.2pt}_{\:p^2=\lmu^2} = 1\;,\nonumber
\ee
where $\gamma_s^{(1)}$ is the one-loop anomalous dimension of the scalar current, which controls the ultra-violet divergence.\footnote{Since we work in the Landau gauge, there is no anomalous dimension for the quark wave-function renormalization constant.}
So, at one loop the renormalization condition becomes:
\be\label{eq:1loopRI}
-z_q^{(1)} + z_s^{(1)} - \gamma_s^{(1)}\n{log}(\HAT{p}^2) + o_s^{(1)} = 0
\ee
It is worthwhile to note the logarithm term of \eqref{eq:1loopRI} just compensates the diverging behaviour of the lattice term $o_s^{(1)}$, so that the whole expression is finite. Hence, the right prescription to get $z_s^{(1)}$ is to subtract $\gamma_s^{(1)}\n{log}(\HAT{p}^2)$ from $o_s^{(1)}$, for each momentum, before taking the continuum limit. Actually, finite volume effects has to be taken into account as explained in \cite{FranzLattice2005}.
With these finite quantities, now, a little algebra (and the previous computed quark wave-function renormalization factors) allows to extract $z_s^{(1)}$.
As order increases the algebra become somewhat more and more involved. In any case, at each order one is always able to extract the renormalization constant, in term of a finite number of log-subtractions.

\end{document}